\documentclass[a4paper,12pt,twoside]{article}

\usepackage{graphicx,natbib}
\usepackage{fancyhdr}
\usepackage{url}
\usepackage{amsmath}
\usepackage{bookmark}
\usepackage{hyperref}
  \hypersetup{
         colorlinks=true,
         linktoc=all,
         linkcolor=cyan,
         urlcolor=blue,
         citecolor=magenta}
\usepackage[titletoc]{appendix}
\usepackage{setspace}

\usepackage[nottoc,numbib]{tocbibind}
\settocbibname{References}

{\normalsize }
\title{Fitting trends in quasar emission and absorption line redshifts}
\author{
Netra K. Subramanian 
\footnote{netra811@gmail.com}~~~~
Prasad Subramanian 
\footnote{Indian Institute of Science Education and Research, Dr Homi Bhabha Road, Pashan, Pune - 411008, India, 
p.subramanian@iiserpune.ac.in, \url{https://www.iiserpune.ac.in/\~p.subramanian}}~~~~ 
Nimisha G. Kantharia  
\footnote{nkprasadnetra@gmail.com, \url{https://sites.google.com/view/nimisha-kantharia}}}
\date{April 2026}

\begin{document}
\maketitle
\tableofcontents\label{toc}

\thispagestyle{empty}

\begin{abstract}
The spectrum of a quasar consists of a few emission lines whose wavelengths are shifted by similar redshifts and numerous absorption lines whose wavelengths are shifted by different redshifts. Hence each quasar is characterised by an emission line redshift and the absorption lines redshifts are all less than the emission line redshift. The distribution of observed absorption line redshifts ($z_{abs}$) with respect to emission line redshift ($z_{em}$) for a large sample of quasars shows a systematic trend as pointed out by \citet{2016arXiv160901593K}. They noticed that increase in $z_{em}$ is accompanied by a monotonic increase in the lowest detected value of $z_{abs}$ and inferred that the emission and absorption lines were all formed in the quasar. 

This study focuses on modeling the systematic trend in the observed $z_{em} \rightarrow z_{abs}$ distribution. We considered the redshift data of absorption lines of singly ionized magnesium (denoted by MgII) and triply ionized carbon (denoted by CIV) for a large sample of quasars. We find that the envelope of data points defining the lowest value of the MgII absorption line redshift (which we denote by $z_{MgIImodel}$) for a given $z_{em}$ satisfies $z_{MgIImodel} = (0.418 \pm 0.008) z_{em} - (0.482 \pm 0.02)$ with an $R^2$ value of 0.99. The model can be used to predict the lowest expected MgII absorption line redshift for any $z_{em}$. We find a similar model for the lowest expected redshift of triply ionized carbon lines for any $z_{em}$ which is $z_{CIVmodel} = 0.845 (\pm 0.0002) z_{em} - 0.153 (\pm 0.0006)$. 

\end{abstract}

\section{Introduction to quasar spectra and redshifts}
\begin{figure}[t]
\centering
\includegraphics[width=13cm]{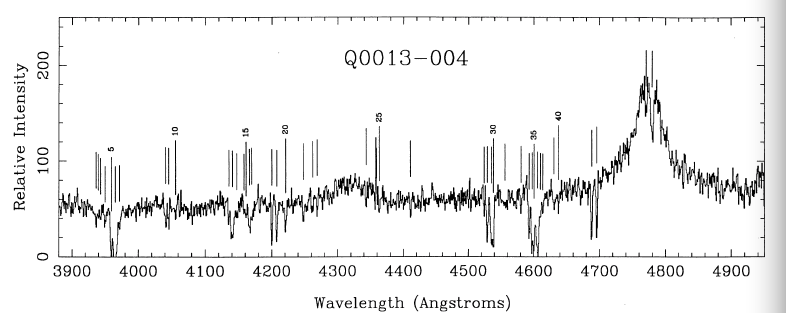}(a)
\includegraphics[width=13cm]{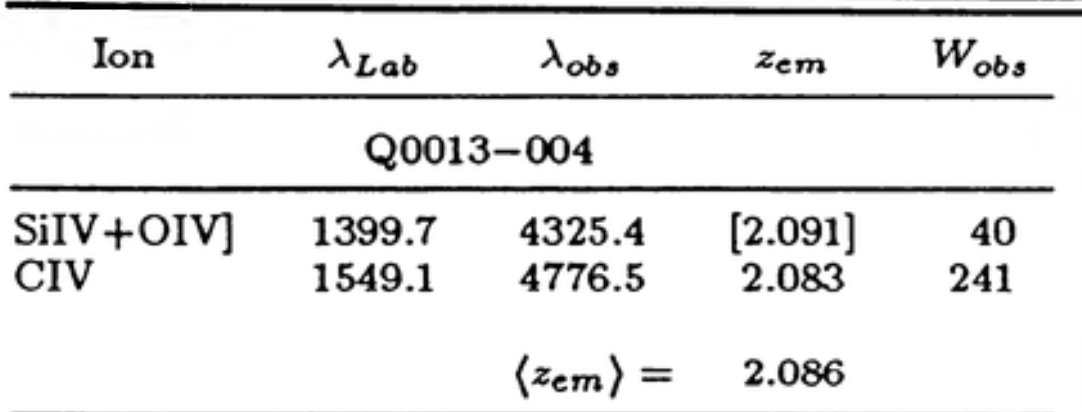}(b)
\caption{\small (a) A typical spectrum recorded between wavelengths of $\sim 3900$ Angstroms and $\sim 5000$Angstroms is shown in this figure taken from \citet{1988ApJS...68..539S}. This spectrum is of quasar Q0013-004. (b) The emission line features are identified and their redshift is listed in this snapshot from \citet{1988ApJS...68..539S}. The emission line redshift of the quasar Q0013-004 is $z_{em}=2.086$.}
\label{fig1}
\end{figure}
\begin{figure}
\centering
\includegraphics[width=13cm]{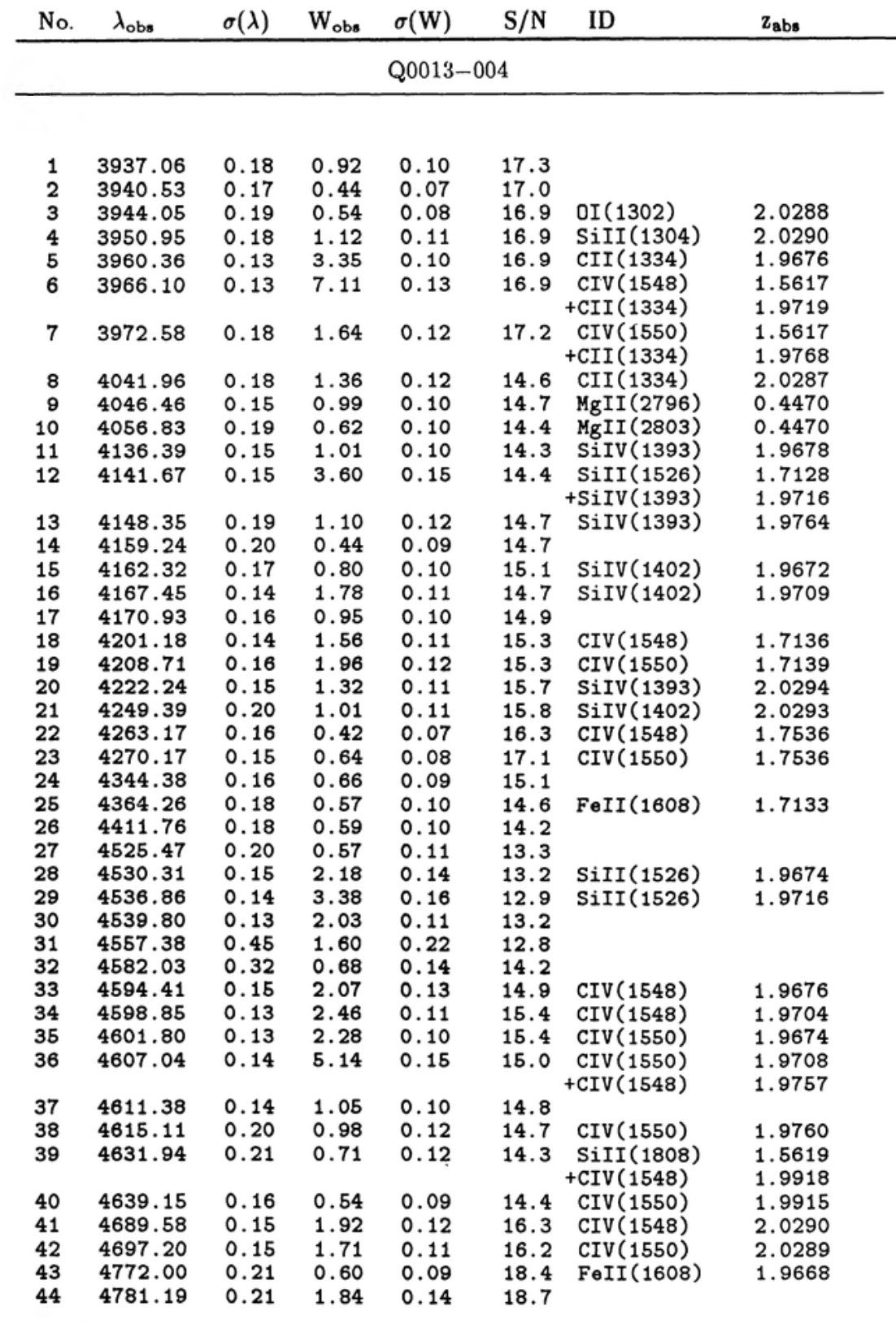}
\caption{\small The absorption features in the spectrum of quasar Q0013-004 (\autoref{fig1}(a)) are listed in this snapshot of table from \citet{1988ApJS...68..539S}. 44 absorption features due to transitions in elements like carbon, iron, silicon, magnesium are identified and the estimated redshifts are listed in the last column of the table. The absorption lines are shifted by redshifts between 2.029 and 0.4470 with the absorption lines of MgII being detected with the lowest redshifts. The emission line redshift of this quasar is  estimated to be 2.086. All absorption line redshifts are lower than the emission line redshift. While emission lines are detected at similar redshifts, absorption lines from the same transition span a range of redshifts. }
\label{fig1a}
\end{figure}

Quasars refer to the supermassive black holes (mass $\ge 10^9$ solar masses) at the centres of some galaxies. They have a compact star-like appearance and often outshine the host galaxy. Quasars are the most luminous active galactic nuclei (AGN). A spectrum recorded in ultraviolet to optical bands by pointing a telescope towards a quasar often detects a few emission lines and numerous absorption lines as illustrated by the spectrum of quasar Q0013-004 (\autoref{fig1}a). The spectral features that are identified are listed in the tables shown in \autoref{fig1}(b) and \autoref{fig1a} taken from \citet{1988ApJS...68..539S}. The spectral lines are due to transitions in different elements in various ionization states. The absorption lines in \autoref{fig1}(a) are attributed to singly ionized carbon (CII), triply ionized silicon (SiIV), singly ionized magnesium (MgII), triply ionized carbon (CIV) and singly ionized silicon (SiII) with the same transition being detected with multiple redshifts. The electromagnetic energy expected to be emitted in these transitions is known and the wavelength/frequency of the emitted electromagnetic wave is referred to as rest wavelength/frequency.  Spectral lines are detected at longer wavelengths instead of the rest wavelengths. This shift in wavelength is known as redshift if the line is displaced to longer wavelengths and is known as blueshift if the line is detected at shorter wavelengths. For electromagnetic waves travelling in vacuum, the speed of light $c = \lambda \nu$ where $\lambda$ refers to wavelength and $\nu$ refers to frequency of the waves. Redshift of the electromagnetic waves can be estimated from wavelength by:
\begin{equation}
\tt z = \frac{\lambda_{obs} - \lambda_{rest}}{\lambda_{rest}}
\label{eqn1}
\end{equation}
where $\lambda_{rest}$ refers to the wavelength of emitted electromagnetic waves i.e. rest wavelength while $\lambda_{obs}$ refers to the wavelength at which the spectral line is observed in the quasar spectrum. This formula is in Doppler convention. The redshift in \autoref{eqn1} can also be obtained from frequency of spectral line from the expression:
\begin{equation}
\tt z = \frac{\nu_{rest} - \nu_{obs}}{\nu_{obs}}
\end{equation}
In quasar spectra, all the observed spectral lines are displaced to a longer wavelength compared to their rest wavelength i.e. are redshifted. In \autoref{fig1}(b), the estimated redshift of two emission lines are listed in the column under the heading $z_{em}$. Since the two values are similar which is typical in quasar spectra, a single emission line redshift is given for a quasar which for Q0013-004 is $z_{em}=2.086$ (see \autoref{fig1}b). In \autoref{fig1a}, the absorption lines in the spectrum of quasar Q0013-004 have been identified. The rest wavelength of the absorption line (in Angstroms) is noted in brackets in column 7 next to the element and the redshift of the absorption lines estimated by using \autoref{eqn1} is listed in the last column under the heading $z_{abs}$. Notice the large range in values of $z_{abs}$ estimated for lines detected in the spectrum of the quasar Q0013-004 which is typical of quasar spectra. 

 The same transition is also detected at several redshifts. 
 For example, there are eight rows with the entry CIV(1550) in column 7 of the table in \autoref{fig1a}. CIV refers to triply ionized carbon and 1550 refers to rest wavelength of the spectral line. These spectral lines are detected in the spectrum of the quasar Q0013-004 at wavelengths displaced 
 from 1550 Angstroms by redshifts of 1.5617, 1.7139, 1.7536, 1.9674, 1.9708, 1.9760, 1.9915, 2.0289 (see \autoref{fig1a}). The origin of the multiple redshifts can either be local to the quasar i.e. all spectral features are formed in the quasar or could be external to the quasar. If local to the quasar then the multiplicity of the observed redshifts might be due to combination of at least two component redshifts of which one component is variable.
Let the two component redshifts be $z_1$, $z_2$ defined in the Doppler convention given in \autoref{eqn1} which sequentially shift the wavelength of a spectral line so that it is detected at a redshift $z$. The aggregate displacement of the spectral line quantified by redshift $z$ is related to components $z_1, z_2$ by
\begin{equation}
\tt (1+z) = (1+z_1)(1+z_2)
\label{eqn1a}
\end{equation}
If we know any two of $z,z_1,z_2$ then the third redshift can be estimated from the above equation. This expression can be obtained by estimating the LHS and RHS separately after substituting the redshift by wavelength from \autoref{eqn1}. It is to be kept in mind that 
$z_1 = \frac{\lambda_{obs,1} - \lambda_{rest}}{\lambda_{rest}}$ and 
$z_2 = \frac{\lambda_{obs} - \lambda_{obs,1}}{\lambda_{obs,1}}$. 

In quasar spectra, the emission line shows the largest redshift as for quasar Q0013-004. Absorption features appear at different redshifts, all lower than the emission line redshift. It is believed that the emission line redshift is cosmological redshift of the quasar and hence is often referred to as $z_{qso}$. It is believed that the shift in wavelength of the spectral lines is due to Doppler effect i.e. due to large relative motion of the emitter and observer. However these remain debatable due to lack of convincing justification.

In this paper, we find model fits to observed trends in redshifts of the emission line and absorption lines detected in quasar spectra. Large datasets available in literature are used for this purpose.

\section{The redshift distribution}
\begin{figure}[t]
\centering
\includegraphics[width=13cm]{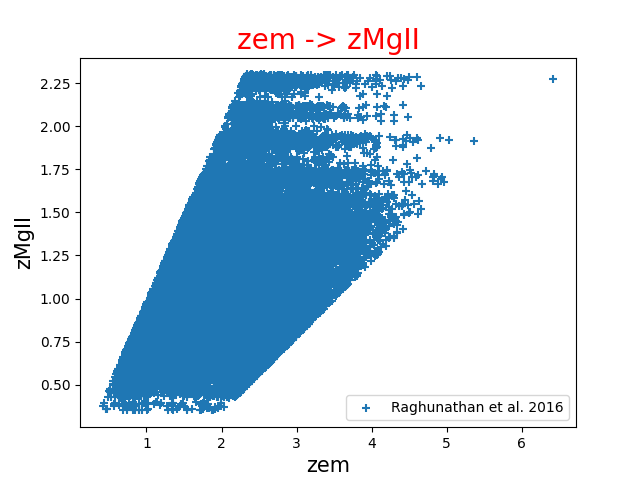}
\caption{\small The redshift distribution of redshift of MgII doublet lines detected in absorption is plotted against the emission line redshift of the quasar for a large sample of quasars. The data have been taken from the catalogue based on SDSS DR12 \citep{2016MNRAS.463.2640R}. Notice the horizontal bounds at $z_{MgII}=0.35$ and $z_{MgII}=2.3$ which defines the range of MgII redshifts in the catalogue. The upper diagonal line is for $z_{MgII}=z_{em}$ while the lower diagonal line is unexpected. Data is sparse in the top right corner of the distribution.}
\label{fig2}
\end{figure}

Let us consider absorption lines of singly ionized magnesium (MgII) detected in the spectra of quasars and refer to their redshift as $z_{MgII}$. These are doublet lines of rest wavelengths 2798 Angstroms and 2803 Angstroms. These lines are displaced by a redshift of 0.4470 in the observed spectrum of Q0013-004 whose $z_{em}=2.086$ (see \autoref{fig1a}). For every quasar in which the doublet lines of MgII are detected, there will be a pair of redshifts ($z_{em}, z_{MgII}$). For quasars in which multiple lines of MgII are detected at different redshifts there will be several such pairs i.e. ($z_{em}, z_{MgII,i}$) where $i$ refers to the multiplicity of MgII doublet features.
When these redshift pairs are plotted for a large sample of quasars, it results in distribution shown in \autoref{fig2}. These data have been taken from the catalogue given by \citet{2016MNRAS.463.2640R}.

The redshift data of absorption lines of the MgII doublet detected in the spectra of a large sample of quasars from the 12th data release (DR12) of Sloan Digital Sky Survey (SDSS) by \citet{2016MNRAS.463.2640R} was used for our analysis. \citet{2016MNRAS.463.2640R} searched for the magnesium doublet line in the spectra of more than 250000 quasars and list 36981 separate MgII doublet features in their catalogue. The redshifts $z_{MgII}$ of these features whose rest wavelengths are 2796 Angstroms and 2803 Angstroms is restricted to the range 0.35 to 2.3 which are the horizontal bounds of the distribution in \autoref{fig2}.  $z_{em}$ of the quasar is on x-axis and $z_{MgII}$ detected in that quasar's spectrum is on y-axis i.e. data from column 5 ($z_{em}$) and column 7 ($z_{MgII}$) from the catalogue published in \citet{2016MNRAS.463.2640R} are plotted in \autoref{fig2}. Each data point represents a pair ($z_{em}, z_{MgII}$). The envelope of the redshift distribution in \autoref{fig2} is trapezoidal with the top right corner of the distribution being determined by a lone datapoint at $z_{em} > 6$.

\subsection{Modelling the trend in $z_{em} \rightarrow z_{MgII}$ distribution }
Let us understand the nature of envelope of the redshift distribution in \autoref{fig2}. The upper and lower horizontal sides of the trapezoidal distribution are due to the detected absorption features of MgII being restricted to the redshift range $0.35 < z_{MgII} < 2.3$. The upper diagonal of the distribution is because $z_{MgII} \le z_{em}$ and is well accounted for by the line $z_{MgII} = z_{em}$. The lower diagonal of the distribution is unexpected and had not been recognized till pointed out by \citet{2016arXiv160901593K}. The trend indicated by the lower diagonal is that an increase in $z_{em}$ is accompanied by a monotonic increase in the lowest $z_{MgII}$ that is detected in any quasar with $z_{em}$. This implies that $z_{MgII}$ and $z_{em}$ are not independent. As pointed out by \citet{2016arXiv160901593K}, the nature of the lower envelope can only be explained if all the emission and absorption features detected in a quasar spectrum arise in the quasar or its host galaxy. If the emission features were formed in the quasar and the absorption features were formed in the medium between us and the quasar then no relation between $z_{em}$ and $z_{abs}$ can be expected. That there is a relation given by the lower diagonal in \autoref{fig2} means $z_{MgII}$ knows what $z_{em}$ is and vice versa which can only be true if they are not independent. In other words, the behaviour of the lower diagonal in \autoref{fig2} firmly establishes the association of absorption lines with the quasar environs. 
\begin{figure}[t]
\centering
    \includegraphics[width=13cm]{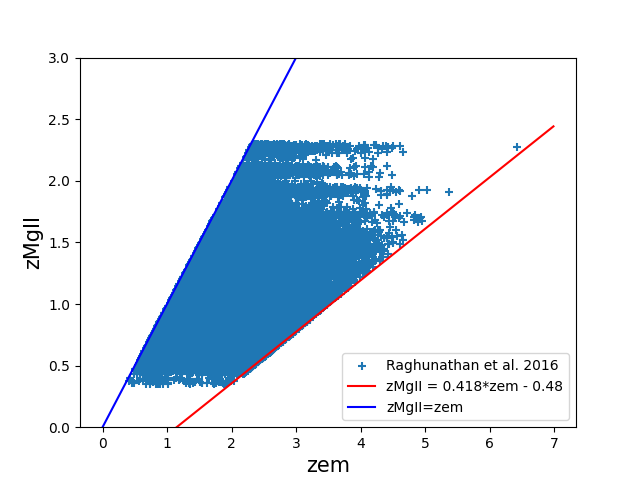}
    \caption{\small The model fitted to the lower envelope of the redshift distribution is plotted in red. The blue line is plotted for x=y. Notice how all the detected redshifts of MgII absorption lines in all quasar spectra lie between the two bounding lines. While the x=y bound is expected since emission line redshifts are generally the highest redshift in a quasar spectrum. However the other bound shown by the red line is unexpected. }
    \label{Mg}
\end{figure}
If the lower side of the envelope of redshift distribution (see \autoref{fig2}) can be fitted then the model can be used to predict the lowest redshift of MgII line that will be detectable for quasars of larger $z_{em}$. 

Since the lowest redshift detected in quasars of redshift $z_{em}$ varied, selecting the lowest $z_{MgII}$ for all quasars resulted in a distribution similar to that in \autoref{fig2} with fewer points. The best way forward was to fit the envelope of the redshift distribution in \autoref{fig2}. 
We implemented this using the Convex Hull algorithm available in Python library {\it scipy.spatial} namely {\it ConvexHull(x,y)} where (x,y) is a 2 dimensional array.
The algorithm gave four points for each side of the envelope i.e. a total of 16 points characterized the four sides of the trapezoidal distribution. Since we are interested in the lower side of the envelope in \autoref{fig2}, we considered only that data from the algorithm.  The best fit to the lower envelope was found to be the straight line given by
\begin{equation}
 z_{MgIImodel} = 0.418 (\pm 0.008) z_{em} - 0.48 (\pm 0.02)   
\label{eqn2}
\end{equation}
 The fit was excellent since $R^2 = 0.99$ and there was no need for a higher order polynomial fit. $z_{MgIImodel}$ refers to the lowest redshift detected for a $z_{em}$. This model is the red line plotted with the data in \autoref{Mg}. The blue line is $z_{MgII}=z_{em}$. As can be seen in the figure, the model is an excellent fit to the lower envelope of the distribution. From \autoref{eqn2}, the lowest redshift at which MgII doublet will be detected in a quasar spectrum for a given emission line redshift $z_{em}$ can be predicted. For example, the lowest redshift of MgII absorption in a quasar of $z_{em}=7$ is $z_{MgIImodel} =2.446$. There will be numerous quasars whose emission line redshift is $z_{em}=7$. These quasar spectra can detect numerous MgII features which will be shifted by redshifts $ 2.446 \le z_{MgII} < 7 $. There will not be a single quasar with a MgII feature displaced by a redshift less than 2.446. 
Let us consider quasars with $z_{em} = 4$. According to the model in \autoref{eqn2}, the lowest redshift of MgII line is 1.19. In other words, no quasar with $z_{em}=4$ shows a MgII feature shifted by a redshift that is less than 1.19 (see \autoref{Mg}). Such a trend appears mysterious but is the kind of observational result which generally will have a unique interpretation. 

 Another way to understand the trend quantified by \autoref{eqn2} is that the difference between $z_{em}$ and $z_{MgII}$ has an upper bound. This upper bound is obtained from the difference between the two redshifts which can be estimated from \autoref{eqn1a} as:
 \begin{equation}
\Delta z = \frac{1+z_{em}}{1+z_{MgIImodel}} - 1
 \end{equation}
 Thus, for the pair $z_{em},z_{MgIImodel} = 7, 2.446$, the difference redshift is $\Delta z=1.32$ while for the pair $z_{em},z_{MgIImodel} = 4, 1.19$, the difference redshift is $\Delta z = 1.28$. Notice that the two difference redshifts are similar although the two $z_{em}$ are very different. 
 
In next two subsections, the model in \autoref{eqn2} is tested on spectral data of quasars of larger $z_{em}, z_{MgII}$ and for absorption lines of other elements.

\subsection{Checking the model on higher redshift data}
\begin{figure}[t]
\centering
\includegraphics[width=13cm]{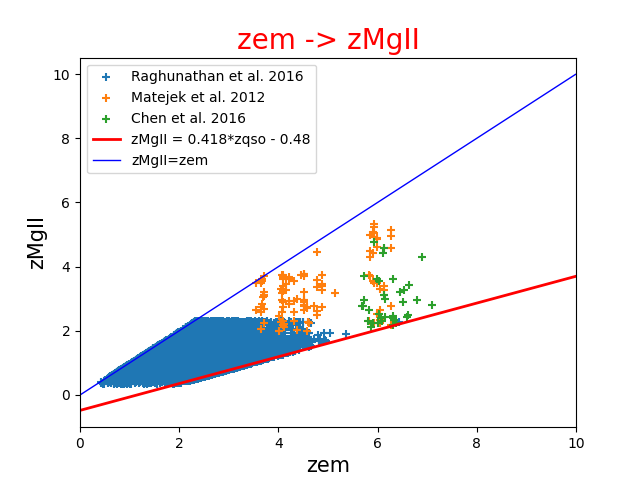}
\caption{\small Redshift data from three databases are plotted along with the model from \autoref{eqn2}. The blue line is for $z_{MgII} = z_{em}$. Notice how all the data lie between these two lines.  }
\label{fig3}
\end{figure}
The model given in \autoref{eqn2} was obtained from MgII redshift data ranging from 0.35 to 2.3 that is shown in \autoref{fig2}. It would be interesting to check if it is applicable to spectral lines of MgII with the lowest detected redshifts being greater than 2.3. In \autoref{fig3}, the data from \citet{2016MNRAS.463.2640R} which was used to obtain the model is plotted and the model of \autoref{eqn2} is shown by the red line. The blue line is $z_{MgII} = z_{em}$. Additionally redshift data from higher emission line quasars (and higher $z_{MgII}$) are also plotted.  The data points in orange are from \citet{2012ApJ...761..112M} while the green data points are from \citet{2017ApJ...850..188C}. Both the data sets lie above the red line as expected from the model. Clearly the model is applicable to MgII absorption lines shifted by redshifts $> 2.3$ in higher emission line redshift quasars. We can infer that the trend in the redshift distribution is followed at least till $z_{em}\sim 7.5$. The model is extrapolated upto $z_{em}=10$ for which $z_{MgIImodel}=3.7$. The difference redshift is $\Delta z = 1.34$. To the best of our knowledge the combination $z_{em}=10, z_{MgIImodel}=3.7$ has not yet been detected. 

\subsection{Trends in redshifts of other absorption lines}
\begin{figure}
\centering
\includegraphics[width=13cm]{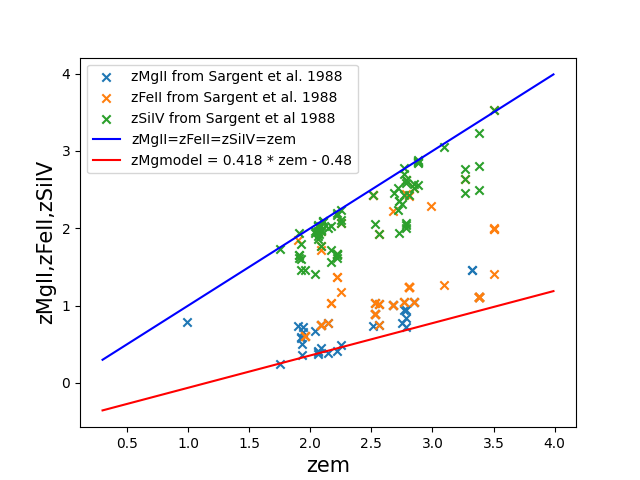}(a)
\includegraphics[width=13cm]{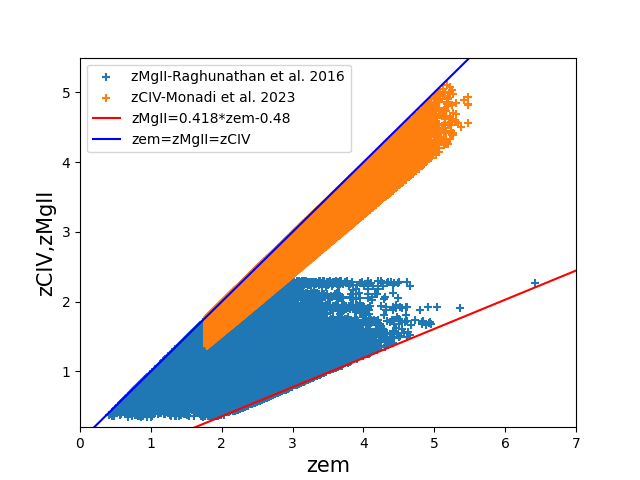}(b)
\caption{\small (a) Redshifts of absorptions lines MgII taken from \citet{1988ApJ...334...22S}, redshifts ofFeII and SiVI taken from \citet{1988ApJS...68..539S} are plotted against the emission line redshift for a small sample of quasars. The red line is the best fit model obtained for $z_{em}\rightarrow z_{MgII}$ distribution. (b) Redshifts of absorption lines of MgII for a large sample of quasars taken from \citet{2016MNRAS.463.2640R} and absorption line redshifts of CIV doublet lines observed for a large sample of quasars taken from \citet{2023MNRAS.526.4557M} are shown. The model from \autoref{eqn2} is shown in red. Notice how all the redshifts lie between the two lines. Notice that the redshift distribution of SiIV lines in (a) resembles that of CIV lines with both lying near the upper diagonal.}
\label{fig4}
\end{figure}
Quasar spectra typically include absorption lines of singly ionized iron (FeII), singly ionized carbon (CII), singly ionized silicon (SiII), triply ionized silicon (SIV), triply ionized carbon (CIV) etc in addition to singly ionized magnesium (MgII). As mentioned earlier, these absorption lines are detected at longer wavelengths than their expected wavelengths with the redshifts spanning a range of values such that all are $\le z_{em}$. Redshifts of absorption lines of MgII (blue points), FeII (orange points) and SiIV (green points) detected in a sample of quasar spectra by \citet{1988ApJS...68..539S,1988ApJ...334...22S} are plotted against $z_{em}$ of the quasar in \autoref{fig4}(a). The red line is the model  (\autoref{eqn2}) while the blue line is for $z_{em}=z_{MgII}=z_{FeII}=z_{SiIV}$. Interestingly all the redshift data lie between these two lines with the MgII redshifts being the lowest redshifts at a given $z_{em}$. The distribution of the redshifts of the different species of lines is distinct. The MgII redshifts lie close to the red line followed by the redshifts of FeII lines while the redshifts of the SiIV lines appear to be the largest for a given $z_{em}$ and lie close to the blue line. The redshifts of triply ionized carbon (CIV) $z_{CIV}$ detected in absorption in quasar spectra for a large sample of quasars are plotted along with $z_{MgII}$ in \autoref{fig4}(b). The CIV lines are a doublet whose rest wavelengths are 1548 Angstroms and 1550 Angstroms. The redshift distributions of the two species show distinct behaviour but systematic trends are seen in both distributions. Interestingly the region in the plot occupied by $z_{CIV}$ is similar to the region occupied by $z_{SiIV}$ in \autoref{fig4}(a). The $z_{CIV}$ data were taken from \citet{2023MNRAS.526.4557M}. The model from \autoref{eqn2} is shown by the red line. 

The systematic trend in the distribution of $z_{em} \rightarrow z_{CIV}$ is the lower side of the envelope of the orange data points in \autoref{fig4}(b). This kind of behaviour by the redshifts is unexpected and means that $z_{em}$ and $z_{CIV}$ are not independent.
\begin{figure}
\centering
    \includegraphics[width=13cm]{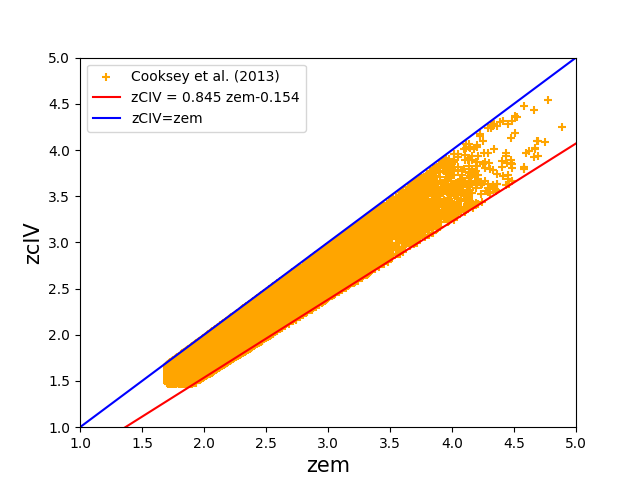}
    \caption{\small The redshift data on CIV doublet lines taken from catalogue of \citet{2013ApJ...763...37C} is plotted. The best model fit to the lower diagonal estimated using data in the range $2 < z_{CIV} < 4$ is $\tt z_{CIVmodel} = 0.845 z_{em} - 0.153$ which is shown by the red line. The blue line is $\tt z_{CIVmodel}=z_{em}$. All the data lie between these two lines.}
    \label{CIV}
\end{figure}
As with the MgII redshifts, we fitted the trend in the CIV redshifts using the data catalogued by \citet{2013ApJ...763...37C}. This catalogue contains redshift data of CIV lines in the redshift range $1.5 < z_{CIV} < 4.5$ extracted from quasar spectra released in SDSS DR7. We used the data between redshifts $2 < z_{CIV} < 4$ for the fitting. As was done with the MgII redshift data, the Convex Hull algorithm in Python was used to characterize the envelope of the distribution. Three sides of the envelope are expected while the lower diagonal is the newly recognized trend. We found that the best linear fit to this side was
\begin{equation}
    \tt z_{CIVmodel} = 0.845 (\pm 0.0002) z_{em} - 0.153 (\pm 0.0006)
\end{equation}
This model is shown by the red line in \autoref{CIV} while the blue line is $z_{CIV}=z_{em}$ and the data points are in orange. Using this model, the lowest value of $z_{CIV}$ for a given $z_{em}$ can be determined. In other words, while absorption lines of CIV with redshift $ > z_{CIVmodel}$ will be detected, no line with redshift $ < z_{CIVmodel}$ is detected at a given $z_{em}$. This is true for a large sample of quasars. All the measured redshifts of CIV absorption lines satisfy $ z_{CIVmodel}< z_{CIV} < z_{em}$.

\section{Summary}
The main results of the paper can be summarised to be:
\begin{itemize}
 \item When the redshifts by which absorption lines detected in a quasar spectrum are displaced $z_{abs}$ from their rest wavelengths are plotted against the quasar emission line redshift $z_{em}$, then a systematic trend is noticed such that increase in $z_{em}$ is accompanied by an increase in the lowest detected $z_{abs}$. 

 \item We find the best fit model for the systematic trend seen in the $z_{em} \rightarrow z_{MgII}$ distribution is $z_{MgIImodel} = 0.418 (\pm 0.008) z_{em} - 0.48 (\pm 0.02)$ with $R^2=0.99$. Here $z_{MgII}$ refers to observed redshifts of singly ionized magnesium lines in absorption while $z_{MgIImodel}$ refers to the lowest $z_{MgII}$ for a $z_{em}$. The observed MgII redshifts obey $z_{MgIImodel}<z_{MgII}<z_{em}$.

 \item The best fit model for the systematic trend in the $z_{em}\rightarrow z_{CIV}$ distribution is $z_{CIVmodel} = 0.845 (\pm 0.0002) z_{em} - 0.153 (\pm 0.0006)$ with $R^2=0.99$. Here $z_{CIV}$ refers to the observed redshifts of triply ionized carbon in absorption while $z_{CIVmodel}$ refers to the lowest $z_{CIV}$ for a $z_{em}$.
 The observed CIV redshifts obey $z_{CIVmodel}<z_{CIV}<z_{em}$.
 \item Similar trends are shown by the redshifts of other absorption lines.
   \end{itemize}

The main implications of these results are:
\begin{enumerate}
  \item The models enable us to estimate the lowest $z_{MgIImodel}$ and $z_{CIVmodel}$ for any $z_{em}$. 
  \item The observed increase in $z_{MgIImodel}$ and $z_{CIVmodel}$ with increasing $z_{em}$ means that the redshifts of the absorption lines and emission lines are not independent. It also implies that there is an upper bound on the difference redshift $\Delta z = \frac{1 + z_{em}}{1+z_{MgIImodel}} - 1$ for magnesium lines and similarly for carbon lines.   
  \item If it was true that emission lines are formed in the quasar and the absorption lines are formed in the intervening medium between us and the quasar then as $z_{em}$ increased, the largest $z_{abs}$ would increase as is observed. But there should be no influence on the lowest $z_{abs}$. In fact, the lowest $z_{abs}$ should be the same for all $z_{em}$. The observed systematic increase in lowest $z_{abs}$ with increasing $z_{em}$ means that in the intervening origin, distant quasars detect only distant intervening media and not local ones. This effectively rules out the intervening medium origin for the absorption lines. 
  
  \item Our findings imply that all spectral lines are formed in the quasar. This, in turn, means that the lowest detected redshift has to be its cosmological redshift. The redshift distributions of emission and absorption lines should be studied against the lowest redshift which is often $z_{MgII}$. 
  
  \item  $z_{em}$ which is the largest redshift in quasar spectrum is not its cosmological redshift.
  
  \item The absorption and emission lines detected in quasar spectra are displaced to longer wavelengths due to two redshift components - (1) a cosmological redshift which is the same for all lines from a quasar and (2) a variable redshift component.

  
\end{enumerate}
There is a significant caveat regarding implications 1 through 6 we have listed above. Since \citet{2016MNRAS.463.2640R} look only for MgII lines that are redward of the Lyman alpha peak in the QSO spectrum, our results might actually reflect the position of the Lyman emission line\footnote{We thank Ian Smail for pointing this out}.

\bibliography{file.bib}

\begin{thebibliography}{}

\bibitem[{Chen} et~al., 2017]{2017ApJ...850..188C}
{Chen}, S.-F.~S., {Simcoe}, R.~A., {Torrey}, P., {Ba{\~n}ados}, E., {Cooksey},
  K., {Cooper}, T., {Furesz}, G., {Matejek}, M., {Miller}, D., {Turner}, M.,
  {Venemans}, B., {Decarli}, R., {Farina}, E.~P., {Mazzucchelli}, C., and
  {Walter}, F. (2017).
\newblock {`Mg II Absorption at $2 < Z < 7$ with Magellan/Fire. III. Full
  Statistics of Absorption toward 100 High-redshift QSOs'}.
\newblock {\em The Astrophysical Journal}, 850(2):188.
\newblock Available from
  \url{https://iopscience.iop.org/article/10.3847/1538-4357/aa9707/pdf}.

\bibitem[{Cooksey} et~al., 2013]{2013ApJ...763...37C}
{Cooksey}, K.~L., {Kao}, M.~M., {Simcoe}, R.~A., {O'Meara}, J.~M., and
  {Prochaska}, J.~X. (2013).
\newblock {`Precious Metals in SDSS Quasar Spectra. I. Tracking the Evolution
  of Strong, 1.5 $<$ z $<$ 4.5 C IV Absorbers with Thousands of Systems'}.
\newblock {\em The Astrophysical Journal}, 763:37.
\newblock Available from
  \url{https://iopscience.iop.org/article/10.1088/0004-637X/763/1/37/pdf}.

\bibitem[{Kantharia}, 2016]{2016arXiv160901593K}
{Kantharia}, N.~G. (2016).
\newblock {`Decoding quasars: gravitationally redshifted spectral lines!'}.
\newblock {\em ArXiv eprint arXiv:1609.01593}.
\newblock Available from
  \url{https://ui.adsabs.harvard.edu/abs/2016arXiv160901593K}. Also look at
  \url{https://sites.google.com/view/nimisha-kantharia/researchpapers} for
  additional relevant documents.

\bibitem[{Matejek} and {Simcoe}, 2012]{2012ApJ...761..112M}
{Matejek}, M.~S. and {Simcoe}, R.~A. (2012).
\newblock {`A Survey of Mg II Absorption at 2 $<$ z $<$ 6 with Magellan/FIRE.
  I. Sample and Evolution of the Mg II Frequency'}.
\newblock {\em The Astrophysical Journal}, 761:112.
\newblock Available from
  \url{https://iopscience.iop.org/article/10.1088/0004-637X/761/2/112}.

\bibitem[{Monadi} et~al., 2023]{2023MNRAS.526.4557M}
{Monadi}, R., {Ho}, M.-F., {Cooksey}, K.~L., and {Bird}, S. (2023).
\newblock {`Machine learning uncovers the universe's hidden gems: A
  comprehensive catalogue of C IV absorption lines in SDSS DR12'}.
\newblock {\em Monthly Notices of the Royal Astronomical Society},
  526(3):4557--4574.
\newblock Available from
  \url{https://academic.oup.com/mnras/article/526/3/4557/7284398?login=false}.

\bibitem[{Raghunathan} et~al., 2016]{2016MNRAS.463.2640R}
{Raghunathan}, S., {Clowes}, R.~G., {Campusano}, L.~E., {S{\"o}chting}, I.~K.,
  {Graham}, M.~J., and {Williger}, G.~M. (2016).
\newblock {`Intervening Mg II absorption systems from the SDSS DR12 quasar
  spectra'}.
\newblock {\em Monthly Notices of the Royal Astronomical Society},
  463(3):2640--2652.
\newblock Available from
  \url{https://academic.oup.com/mnras/article/463/3/2640/2646550?login=false}.

\bibitem[{Sargent} et~al., 1988a]{1988ApJS...68..539S}
{Sargent}, W.~L.~W., {Boksenberg}, A., and {Steidel}, C.~C. (1988a).
\newblock {`C IV absorption in a new sample of 55 QSOs - Evolution and
  clustering of the heavy-element absorption redshifts'}.
\newblock {\em The Astrophysical Journal Supplement Series}, 68:539--641.
\newblock Available from
  \url{https://articles.adsabs.harvard.edu/pdf/1988ApJS...68..539S}.

\bibitem[{Sargent} et~al., 1988b]{1988ApJ...334...22S}
{Sargent}, W.~L.~W., {Steidel}, C.~C., and {Boksenberg}, A. (1988b).
\newblock {`MG II absorption in the spectra of high and low redshift QSOs'}.
\newblock {\em The Astrophysical Journal}, 334:22--33.
\newblock Available from
  \url{https://articles.adsabs.harvard.edu/pdf/1988ApJ...334...22S}.

\end{thebibliography}

\end{document}